\documentclass[conference,letterpaper]{IEEEtran}

\usepackage{etex}

\def\papertitle{Unrolled and Pipelined Decoders based on Look-Up Tables for Polar Codes}

\usepackage[draft,pdfborder={0 0 0}]{hyperref}

\usepackage{graphicx}
\usepackage[cmex10]{amsmath}
\usepackage{amsthm,amsfonts,amssymb}
\usepackage{bm} 
\usepackage[caption=false,font=footnotesize]{subfig}
\usepackage{booktabs}
\usepackage{cite}
\usepackage{tikz,pgfplots}
\usetikzlibrary{shapes,positioning,arrows,decorations.markings,fit,patterns,arrows.meta,calc,mux,circuits.logic.US,spy}
\pgfplotsset{compat=1.18}

\usepackage{xcolor}
\usepackage{balance}
\usepackage{xfrac} 

\usepackage{glossaries} 

\newcommand{\sgn}[1]{\text{sgn}(#1)}
\newcommand{\mvec}[1]{\bm{#1}}

\input{figures/register.tex} 
\input{figures/register_dotted.tex}

\begin{document}

\bstctlcite{IEEEexample:BSTcontrol}

\newacronym{sc}{SC}{successive-cancellation}
\newacronym{ssc}{SSC}{simplified successive-cancellation}
\newacronym{scl}{SCL}{successive-cancellation list}
\newacronym{ca-scl}{CA-SCL}{CRC-aided \gls{scl}}
\newacronym[plural=CCs,firstplural=clock cycles (CCs)]{cc}{CC}{clock cycle}
\newacronym{llr}{LLR}{log-likelihood ratio}
\newacronym{lut}{LUT}{look-up table}
\newacronym{bpsk}{BPSK}{binary phase-shift keying}
\newacronym{awgn}{AWGN}{additive white Gaussian noise}
\newacronym{fer}{FER}{frame-error rate}
\newacronym{ber}{BER}{bit-error rate}
\newacronym{par}{PAR}{place-and-route}
\newacronym{ib}{IB}{information bottleneck}
\newacronym{msb}{MSB}{most-significant bit}

\title{\papertitle}

\author{\IEEEauthorblockN{Pascal Giard\IEEEauthorrefmark{1}, Syed Aizaz Ali Shah\IEEEauthorrefmark{2}, Alexios Balatsoukas-Stimming\IEEEauthorrefmark{4}, Maximilian Stark\IEEEauthorrefmark{2}, and Gerhard Bauch\IEEEauthorrefmark{2}}
  \IEEEauthorblockA{\IEEEauthorrefmark{1}LaCIME, \'Ecole de technologie sup\'erieure, Montr\'eal, Qu\'ebec, Canada.}
  \IEEEauthorblockA{\IEEEauthorrefmark{2}Institute of Communications, Hamburg University of Technology, Germany.}
  \IEEEauthorblockA{\IEEEauthorrefmark{4}Department of Electrical Engineering, Eindhoven University of Technology, Eindhoven, The Netherlands.}
  Email: pascal.giard@etsmtl.ca, aizaz.shah@tuhh.de, a.k.balatsoukas.stimming@tue.nl, \{maximilian.stark, bauch\}@tuhh.de}%

\maketitle

\begin{abstract} Unrolling a decoding algorithm allows to achieve extremely high throughput at the cost of increased area. \Glspl{lut} can be used to replace functions otherwise implemented as circuits. In this work, we show the impact of replacing blocks of logic by carefully crafted \glspl{lut} in unrolled decoders for polar codes. We show that using \glspl{lut} to improve key performance metrics (e.g., area, throughput, latency) may turn out more challenging than expected. We present three variants of \gls{lut}-based decoders and describe their inner workings as well as circuits in detail. The \gls{lut}-based decoders are compared against a regular unrolled decoder, employing fixed-point representations for numbers, with a comparable error-correction performance. A short systematic polar code is used as an illustration. All resulting unrolled decoders are shown to be capable of an information throughput of little under 10\,Gbps in a 28\,nm FD-SOI technology clocked in the vicinity of 1.4\,GHz to 1.5\,GHz. The best variant of our \gls{lut}-based decoders is shown to reduce the area requirements by 23\% compared to the regular unrolled decoder while retaining a comparable error-correction performance.
\end{abstract}

\glsresetall

\section{Introduction}\label{sec:intro}
Unrolled decoders are known for their extremely high throughput \cite{Schlafer2013,Balatsoukas-Stimming2015,Giard_TCASI_2016,Giard_JETCAS_2017}. In particular, they offer at least one order of magnitude improvement in throughput with respect to standard decoders at the cost of larger area requirements. While this unrolling technique has been applied to \gls{sc}-based polar decoders before, e.g., \cite{Giard_TCASI_2016,Giard_JETCAS_2017}, it has not yet been combined with \gls{lut}-based decoding that has the potential to reduce the required quantization bit-width and, hence, the area and power consumption of the decoder.

\subsubsection*{Contributions}
In this paper, we describe the design and implementation of unrolled and pipelined \gls{lut}-based hardware decoders for polar codes. We present three different variants and provide results for all three, along with results for a regular fixed-point decoder, illustrating the challenges of realizing the \gls{lut}-based decoders in hardware. In the end, we show that, even for a short $(128, 64)$ polar code, a \gls{lut}-based decoder can reduce the area requirements by 23\% while matching the error-correction performance and exceeding the throughput of a decoder using a standard fixed-point representation.

\subsubsection*{Outline} 
The remainder of this paper starts with \autoref{sec:bg} that provides the necessary background, consisting of a brief review of polar codes and an introduction to the \gls{sc} and \gls{ssc} decoding algorithms. Moreover, the concept of unrolled and pipelined hardware architectures is presented as well as that of using \glspl{lut} to implement functions. \autoref{sec:gen} describes our adaptation of the fully-unrolled and pipelined hardware architecture to \gls{lut}-based decoding. In particular, the generation of the \glspl{lut}, the architectures, and the decoders are discussed. \autoref{sec:impl} discusses implementation details and provides post-synthesis ASIC area and timing results using the 28\,nm FD-SOI CMOS technology from ST Microelectronics. 
Finally, \autoref{sec:conclusion} concludes this paper.

\section{Background}\label{sec:bg}
\subsection{Encoding of Polar Codes}\label{sec:bg:enc}
In matrix form, a polar code of length $N$ can be obtained as $\mvec{x} {=} \mvec{u}\mvec{F}^{\otimes n}$, where $\mvec{F} {=} \left[ \begin{smallmatrix} 1 & 0 \\ 1 & 1 \end{smallmatrix} \right]$, $n \triangleq \log _2 N$, $\mvec{u}$ is the vector of bits to be encoded, and $\mvec{F}^{\otimes n}$ is the $n^{\text{th}}$ Kronecker product of $\mvec{F}$ and $\mvec{F}^{\otimes 1}{=}\mvec{F}$. To obtain an $(N,\,k)$ polar code of rate $R=\sfrac{k}{N}$, the $k$ most-reliable bit locations in $\mvec{u}$ are used to hold the information bits while the other $N{-}k$ bits, called frozen bits, are set to a predetermined value (usually 0). The bit-location reliabilities depend on the channel type and condition.

The encoding process can also be represented as a graph like that of Fig.\,\ref{fig:pc8}, where $\oplus$ represents modulo-2 addition (XOR). In that representation, a codeword is generated by setting the frozen-bit locations (the $u_0$ to $u_2$ in light gray) to 0 and the information-bit locations (the $u_3$ to $u_7$ in black) to the message to be encoded, and by propagating the data through the graph, from left to right. As described in \cite{Sarkis_TCOMM_2015}, systematic encoding can be carried out by feeding the output values ($x_0$ to $x_7$) into the left-hand side, resetting the frozen-bit locations to 0, and propagating the data through the graph again.

\subsection{Successive-Cancellation Decoding and Simplified Successive-Cancellation Decoding}\label{sec:bg:sc}
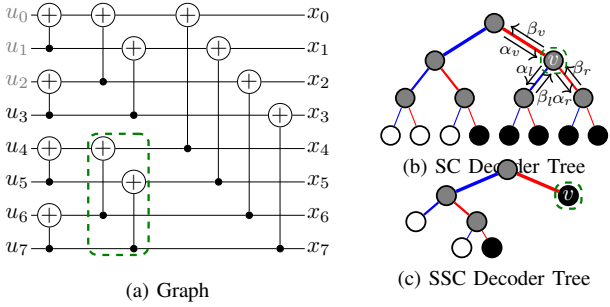
\begin{figure}[t]
  \begin{minipage}{0.6\columnwidth}
    \centering
    \subfloat[Graph]{\label{fig:pc8}\hspace{-15pt}\resizebox{\columnwidth}{!}{\newcommand{\ubit}[1]{$u_{#1}$}
\newcommand{\fbit}[1]{\color{gray}$u_{#1}$}
\newcommand{\ucw}[1]{$x_{#1}$}
\newcommand{\fcw}[1]{\color{gray}$x_{#1}$}
\newcommand{\ub}[1]{$#1$}
\newcommand{\fb}[1]{\color{gray}$#1$}

\begin{tikzpicture}

\usetikzlibrary{shapes,positioning,arrows,decorations.markings,fit}

\definecolor{varnode_fill}{RGB}{0,0,0}
\definecolor{chknode_fill}{RGB}{255,255,255}

\tikzset{
  chknode/.style={draw,fill=chknode_fill,circle,minimum size=0.3cm, inner sep=0},
  varnode/.style={draw,fill=varnode_fill,circle,minimum size=0.1cm, inner sep=0},
  channel/.style={draw,fill=white,rectangle},
  sep/.style={rectangle,minimum width=0.25cm, inner sep=0},
  empty/.style={rectangle, inner sep=0},
  bit/.style={circle, inner sep = 0}
}

\tikzset{green dotted/.style={draw=green!50!black, line width=1pt,
    dash pattern=on 3pt off 3pt,
    inner sep=0.4mm, rectangle, rounded corners}};

\matrix[row sep=1mm, column sep=1mm] {
  \node[bit] (n0s0) {\fb{u_0}}; & \node[chknode] (n0s1) {$+$}; & \node[sep] (s10) {}; & \node[chknode] (n0s2) {$+$}; & \node[empty] {};              & \node[sep] (s20) {}; & \node[chknode] (n0s3) {$+$}; & \node[empty] {}; & \node[empty] {}; & \node[empty] {}; && \node[bit] (xn0s4) {\ub{x_0}};\\
  \node[bit] (n1s0) {\fb{u_1}}; & \node[varnode] (n1s1) {};    & \node[sep] (s11) {}; &                              & \node[chknode] (n1s2) {$+$};  & \node[sep] (s21) {}; & \node[empty] {};             & \node[chknode] (n1s3) {$+$}; & \node[empty] {}; & \node[empty] {}; && \node[bit] (xn1s4) {\ub{x_1}};\\
  \node[bit] (n2s0) {\fb{u_2}}; & \node[chknode] (n2s1) {$+$}; & \node[sep] (s12) {}; & \node[varnode] (n2s2) {};    & \node[empty] {};              & \node[sep] (s22) {}; & \node[empty] {};             & \node[empty] {}; & \node[chknode] (n2s3) {$+$}; & \node[empty] {}; && \node[bit] (xn2s4) {\ub{x_2}};\\

  \node[bit] (n3s0) {\ub{u_3}}; & \node[varnode] (n3s1) {};    & \node[sep] (s13) {}; & \node[empty] {};             & \node[varnode] (n3s2) {};     & \node[sep] (s23) {}; & \node[empty] {};             & \node[empty] {}; & \node[empty] {}; & \node[chknode] (n3s3) {$+$}; && \node[bit] (xn3s4) {\ub{x_3}};\\

  \node[bit] (n4s0) {\ub{u_4}}; & \node[chknode] (n4s1) {$+$}; & \node[sep] (s14) {}; & \node[chknode] (n4s2) {$+$}; & \node[empty] {};              & \node[sep] (s24) {}; & \node[varnode] (n4s3) {};    & \node[empty] {}; & \node[empty] {}; & \node[empty] {}; && \node[bit] (xn4s4) {\ub{x_4}};\\
  \node[bit] (n5s0) {\ub{u_5}}; & \node[varnode] (n5s1) {};    & \node[sep] (s15) {}; &                              & \node[chknode] (n5s2) {$+$};  & \node[sep] (s25) {}; & \node[empty] {};             & \node[varnode] (n5s3) {}; & \node[empty] {}; &  \node[empty] {}; && \node[bit] (xn5s4) {\ub{x_5}};\\
  \node[bit] (n6s0) {\ub{u_6}}; & \node[chknode] (n6s1) {$+$}; & \node[sep] (s16) {}; & \node[varnode] (n6s2) {};    & \node[empty] {};              & \node[sep] (s26) {}; & \node[empty] {};             & \node[empty] {}; & \node[varnode] (n6s3) {}; &  \node[empty] {}; && \node[bit] (xn6s4) {\ub{x_6}};\\
  
  \node[bit] (n7s0) {\ub{u_7}}; & \node[varnode] (n7s1) {};    & \node[sep] (s17) {}; &                              & \node[varnode] (n7s2) {};  & \node[sep] (s27) {}; & \node[empty] {};             & \node[empty] {}; & \node[empty] {}; &  \node[varnode] (n7s3) {}; && \node[bit] (xn7s4) {\ub{x_7}};\\
};
\path[-] (n0s0) edge (n0s1) (n0s1) edge (n0s2) (n0s2) edge (n0s3) (n0s3) edge (xn0s4);
\path[-] (n1s0) edge (n1s1) (n1s1) edge (n1s2) (n1s2) edge (n1s3) (n1s3) edge (xn1s4);
\path[-] (n2s0) edge (n2s1) (n2s1) edge (n2s2) (n2s2) edge (n2s3) (n2s3) edge (xn2s4);
\path[-] (n3s0) edge (n3s1) (n3s1) edge (n3s2) (n3s2) edge (n3s3) (n3s3) edge (xn3s4);
\path[-] (n4s0) edge (n4s1) (n4s1) edge (n4s2) (n4s2) edge (n4s3) (n4s3) edge (xn4s4);
\path[-] (n5s0) edge (n5s1) (n5s1) edge (n5s2) (n5s2) edge (n5s3) (n5s3) edge (xn5s4);
\path[-] (n6s0) edge (n6s1) (n6s1) edge (n6s2) (n6s2) edge (n6s3) (n6s3) edge (xn6s4);
\path[-] (n7s0) edge (n7s1) (n7s1) edge (n7s2) (n7s2) edge (n7s3) (n7s3) edge (xn7s4);

\path[-] (n0s1) edge (n1s1);
\path[-] (n2s1) edge (n3s1);
\path[-] (n4s1) edge (n5s1);
\path[-] (n6s1) edge (n7s1);

\path[-] (n0s2) edge (n2s2);
\path[-] (n1s2) edge (n3s2);
\path[-] (n4s2) edge (n6s2);
\path[-] (n5s2) edge (n7s2);

\path[-] (n0s3) edge (n4s3);
\path[-] (n1s3) edge (n5s3);
\path[-] (n2s3) edge (n6s3);
\path[-] (n3s3) edge (n7s3);

\node (g_n1s2) [green dotted, fit = (n4s2) (n5s2) (n6s2) (n7s2)] {};

\end{tikzpicture}}}
  \end{minipage}%
  \begin{minipage}{0.38\columnwidth}
    \centering
    \subfloat[SC Decoder Tree]{\label{fig:sc-tree}\resizebox{0.9\columnwidth}{!}{\rotatebox{90}{\begin{tikzpicture}[baseline = (0_7.center),
        level/.style={level distance = 6mm},
        level 1/.style={sibling distance=19mm, edge from parent/.style={draw,blue,line width=1.5pt}},
        level 2/.style={sibling distance=9.5mm, edge from parent/.style={draw,blue,line width=1pt}},
        level 3/.style={sibling distance=4.7mm, edge from parent/.style={draw,blue,line width=0.5pt}},
        ]

\tikzset{
frozen/.style={thick,draw=black,fill=white,minimum size=3mm,circle, inner sep=0},
fullspace/.style={thick,draw=black,fill=black,minimum size=3mm,circle, inner sep = 0},
mixed/.style={thick,draw=black,fill=gray,minimum size=3mm,circle, inner sep = 0},
phantom/.style={draw=white,fill=white,minimum size=3mm,circle, inner sep = 0},
}

\tikzset{
parallel segment/.style={
   segment distance/.store in=\segDistance,
   segment pos/.store in=\segPos,
   segment length/.store in=\segLength,
   to path={
   ($(\tikztostart)!\segPos!(\tikztotarget)!\segLength/2!(\tikztostart)!\segDistance!90:(\tikztotarget)$) -- 
   ($(\tikztostart)!\segPos!(\tikztotarget)!\segLength/2!(\tikztotarget)!\segDistance!-90:(\tikztostart)$)  \tikztonodes
   }, 
   segment pos=.5,
   segment length=2.5ex,
   segment distance=-1mm,
},
}

\tikzset{green dotted/.style={draw=green!50!black, line width=0.75pt,
    dash pattern=on 3pt off 3pt,
    inner sep=0.4mm, rectangle, rounded corners}};

\node[mixed] (p){} [grow=left]
	child {node[mixed] (2_0){}
		child {node[mixed] (1_0){}
			child {node[frozen] (a0_0){}
			}
			child {node[frozen] (a0_1){} edge from parent[red]
			}
		}
		child {node[mixed] (1_2){} edge from parent[red]
			child {node[frozen] (0_2){}
			}
			child {node[fullspace] (0_3){} edge from parent[red]
			}
		}
	}
	child {node[mixed] (v){\rotatebox{-90}{\textcolor{white}{$v$}}} edge from parent[red]
		child {node[mixed] (cl){}
			child {node[fullspace] (0_4){}
			}
			child {node[fullspace] (0_5){} edge from parent[red]
			}
		}
		child {node[mixed] (cr){} edge from parent[red]
			child {node[fullspace] (0_6){}
			}
			child {node[fullspace] (0_7){} edge from parent[red]
			}
		}
	}
;

\draw[->,line width=0.65pt] (p) to[parallel segment,segment length=4ex] node[above left=-2.0mm] {\rotatebox{-90}{\footnotesize $\alpha_v$}} (v);
\draw[->,line width=0.65pt] (v) to[parallel segment,segment length=4ex] node[below right=-2.0mm] {\rotatebox{-90}{\footnotesize $\beta_v$}} (p);

\draw[->,line width=0.65pt] (cl) to[parallel segment] (v) {};
\node at ($(cl.south)-(0,0.18)$) {\rotatebox{-90}{\footnotesize $\beta_l$}};
\draw[->,line width=0.65pt] (v) to[parallel segment] node[above right=-2.0mm] {\rotatebox{-90}{\footnotesize $\alpha_l$}} (cl);

\draw[->,line width=0.65pt] (v) to[parallel segment] (cr) {};
\node at ($(cr.north)+(-0.04,0.16)$) {\rotatebox{-90}{\footnotesize $\alpha_r$}};
\draw[->,line width=0.65pt] (cr) to[parallel segment] node[below right=-2.0mm] {\rotatebox{-90}{\footnotesize $\beta_r$}} (v);

\node (g_concat) [green dotted, fit = (v)] {};

\end{tikzpicture}}}}
    \vspace{-11pt}
    \subfloat[SSC Decoder Tree]{\makebox[\columnwidth][c]{\label{fig:ssc-tree}\resizebox{0.7\columnwidth}{!}{\rotatebox{90}{\begin{tikzpicture}[baseline = (v.center),
        level/.style={level distance = 4.0625mm},
        level 1/.style={sibling distance=19mm, edge from parent/.style={draw,blue,line width=1.5pt}},
        level 2/.style={sibling distance=9.5mm, edge from parent/.style={draw,blue,line width=1pt}},
        level 3/.style={sibling distance=4.7mm, edge from parent/.style={draw,blue,line width=0.5pt}},
        ]

\tikzset{
frozen/.style={thick,draw=black,fill=white,minimum size=3mm,circle, inner sep=0},
fullspace/.style={thick,draw=black,fill=black,minimum size=3mm,circle, inner sep = 0},
mixed/.style={thick,draw=black,fill=gray,minimum size=3mm,circle, inner sep = 0},
phantom/.style={draw=white,fill=white,minimum size=3mm,circle, inner sep = 0},
}

\tikzset{green dotted/.style={draw=green!50!black, line width=1pt,
    dash pattern=on 3pt off 3pt,
    inner sep=0.4mm, rectangle, rounded corners}};

\node[mixed] (p){} [grow=left]
	child {node[mixed] (2_0){}
		child {node[frozen] (1_0){}
		}
		child {node[mixed] (1_2){} edge from parent[red]
			child {node[frozen] (0_2){}
			}
			child {node[fullspace] (0_3){} edge from parent[red]
			}
		}
	}
	child {node[fullspace] (v){\rotatebox{-90}{\textcolor{white}{$v$}}} edge from parent[red]
	}
;

\node (g_concat) [green dotted, fit = (v)] {};

\end{tikzpicture}}}}}
  \end{minipage}
  \caption{Graph and decoder-tree representations of an $(8,\,5)$ polar code.}
  \label{fig:pc_8_5}
\end{figure}

The \gls{sc} decoding algorithm was proposed in the seminal work that introduced polar codes~\cite{ArikanFirst}. Illustrating its execution using a decoder-tree representation, it proceeds by visiting the tree---e.g., Fig.~\ref{fig:sc-tree}---sequentially, from top to bottom, from left to right, successively estimating $\bm{\hat{u}}$ at the leaf nodes, from the noisy channel values. Visiting a left edge (blue) on this representation, the \gls{sc} algorithm can calculate the soft-input \glspl{llr} $\alpha_l$ to the child node with the min-sum approximation \cite{Leroux2011}
\begin{equation}\label{eqn:sc:f}
\alpha_l[i] = \sgn{\alpha_v[i]\alpha_v[i + \sfrac{N_v}{2}]} \min(|\alpha_v[i]|, |\alpha_v[i + \sfrac{N_v}{2}]|),
\end{equation}
where $\alpha_v$ and $N_v$ are respectively the \glspl{llr} and node length from the parent node, and $\sgn{x}$ returns $-1$ when $x<0$, $+1$ otherwise. At the root node, the channel \glspl{llr} are used. Once a leaf node is reached, a bit $\hat{u}_i$ (for a non-systematic polar code) is estimated as
\begin{equation}\label{eqn:sc:estu}
\hat{u}_i = \begin{cases}
0\text{,} & \text{when } \alpha_v \geq 0~\text{or}~i \in \mathcal{F};\\
1\text{,} & \text{otherwise,}
\end{cases}
\end{equation}
where $\mathcal{F}$ is the set of frozen-bit indices. For a systematic polar code under \gls{sc} decoding, the estimated-bit vector can be obtained at the end of the decoding process by calculating $\mvec{\hat{u}}\mvec{F}^{\otimes n}$ or its equivalent.

Visiting a right edge (red), the \glspl{llr} $\alpha_r$ to the child node can be calculated \cite{Leroux2011} as
\begin{equation}\label{eqn:sc:g}
\alpha_r[i] = \begin{cases}
\alpha_v[i + \sfrac{N_v}{2}] + \alpha_v[i]\text{,} & \text{when } \beta_l[i] = 0;\\
\alpha_v[i + \sfrac{N_v}{2}] - \alpha_v[i]\text{,} & \text{otherwise},
\end{cases}
\end{equation}
where $\beta_l$ is the bit-estimate vector generated by the left sibling in the decoder-tree. If the left sibling is a leaf node, its estimated-bit value $\hat{u}_i$ is used as the $\beta_l$. Otherwise, the estimated-bit vector $\beta_v$ at a node $v$ is calculated as
\begin{equation}\label{eqn:sc:combine}
\beta_v[i] =
  \begin{cases}
    \beta_l[i]\oplus \beta_r[i], & \text{when}~i < \sfrac{N_v}{2}\\
    \beta_r[i+\sfrac{N_v}{2}], & \text{otherwise},
  \end{cases}
\end{equation}
where $\beta_l$ and $\beta_r$ are the bit-estimate vectors from the left- and right-child nodes, respectively.

The \gls{ssc} algorithm is a variant that exploits the fact that subtrees solely composed of either frozen (rate-0 codes) or information nodes (rate-1 codes) do not need to be fully traversed \cite{Alamdar-Yazdi2011}. 
Fig.\,\ref{fig:ssc-tree} shows a decoder tree for the $(8,\,5)$ polar code of Fig.\,\ref{fig:pc8}, where the \gls{ssc} algorithm is applied, i.e., the tree of Fig.\,\ref{fig:sc-tree} is pruned by recognizing that part of its left-hand-side subtree is a rate-0 code and that the right-hand-side subtree is a rate-1 code.
In Fig.\,\ref{fig:ssc-tree}, the former subtree is replaced by a white node and the latter with a black node. Both cases are direct applications of \eqref{eqn:sc:estu}, i.e., the estimated bit vector for a rate-0 code is always the all-zero vector and that of a rate-1 code is composed of hard decisions on the \glspl{llr} as it does not contain any redundancy.

\subsection{Unrolled and Pipelined Hardware Architectures}\label{sec:bg:unrolled}
\begin{figure}[t]
  \centering
  \resizebox{\columnwidth}{!}{\begin{tikzpicture}[font=\scriptsize,inner sep=1pt, minimum width=1.2em]

  \definecolor{deepgreen}{RGB}{8, 130, 25}

  \tikzset{
    branch/.style={fill,shape=circle,minimum size=3pt,inner sep=0pt},
    block/.style={draw, rectangle, minimum height=2em},
    comb/.style={draw, rectangle, minimum height=6em}
  }

  \node (ac) at (0.2,0) {$\alpha_c$};
  \node at (0.2, 0.63) (cc) {CC};

  \node[shape=reg] at ($(ac)+(0.65,-0.195)$) (REGc0) {$\alpha_c$};

  \node[shape=regdot] at ($(REGc0)+(1.3,0)$) (REGc1) {$\alpha_c$};
  \node[block] at ($(REGc0)+(0.7,-1.0)$) (F1) {$f$};
  \node[shape=reg] at ($(F1)+(0.6,-0.2)$) (REGf1) {$\alpha_1$};

  \node[block,anchor=west] at ($(REGf1.Q)+(0.15,0)$) (G0R1) {$g_{0R}$};
  \node[shape=reg,anchor=west] at ($(G0R1.east)+(0.15,-0.2)$) (REGg0r1) {$\alpha_2$};
  \node[shape=reg] at (REGc1 -| REGg0r1) (REGc2) {$\alpha_c$};

  \node[block,anchor=west] at ($(REGg0r1.Q)+(0.15,0)$) (G0R2) {$g_{0R}$};
  \node[block,anchor=west] at ($(G0R2.east)+(0.15,0)$) (I1) {$I$};
  \node[shape=reg,anchor=west] at ($(I1.east)+(0.15,-0.2)$) (REGi1) {$\beta_1$};
  \node[shape=regdot] at (REGc2 -| REGi1) (REGc3) {$\alpha_c$};

  \node[block,anchor=west] at ($(REGi1.Q)+(0.15,0)$) (C0R1) {$C_{0R}$};
  \node[block] at ($(REGc3.Q -| C0R1.east)+(0.5,0)$) (G1) {$g$};
  \node[block,anchor=west] at ($(G1.east)+(0.15,0)$) (I2) {$I$};

  \node[shape=reg,anchor=west] at ($(I2.east)+(0.15,-0.2)$) (REGi2) {$\beta_2$};
  \node[shape=reg] at ($(C0R1 -| REGi2)+(0,-0.2)$) (REGc0r1) {$\beta_3$};

  \node[comb] at ($(REGi2.east)!0.5!(REGc0r1.east)+(0.4,0)$) (C1) {$C$};
  \node[shape=reg] at ($(C1 |- REGi2)+(0.6,0)$) (REGcomb1) {$\beta_c$};

  \node[anchor=west] (bc) at ($(REGcomb1.east)+(0.15,0.195)$) {$\beta_c$};
  \node at (cc -| bc) {6};

  \draw[-] (ac.east) -- (REGc0.D);

  \draw[dotted] (cc.north -| REGc0.north) -- (REGc0.north) (REGc0.south) -- ($(REGc0.south |- REGf1.south)+(0, -0.2)$);

  \draw[-] (REGc0.Q) -- (REGc1.D);
  \draw[-] ($(REGc0.Q)!0.5!(REGc0.Q-|F1.west)$) node[branch] {} |- (F1);
  \draw[-] (F1.east |- REGf1.D) -- (REGf1.D);

  \draw[dotted] (cc.north -| REGc1.north) -- (REGc1.north) (REGc1.south) -- (REGf1.north) (REGf1.south) -- ($(REGc1.south |- REGf1.south)+(0, -0.2)$);

  \draw[-] (REGc1.Q) -- (REGc2.D);
  \draw[-] (REGf1.Q) -- (G0R1.west) (G0R1.east |- REGg0r1.D) -- (REGg0r1.D);

  \draw[dotted] (cc.north -| REGc2.north) -- (REGc2.north) (REGc2.south) -- (REGg0r1.north) (REGg0r1.south) -- ($(REGc2.south |- REGf1.south)+(0, -0.2)$);

  \draw[-] (REGc2.Q) -- (REGc3.D);
  \draw[-] (REGg0r1.Q) -- (G0R2.west) (G0R2.east) -- (I1.west) (I1.east |- REGi1.D) -- (REGi1.D);

  \draw[dotted] (cc.north -| REGc3.north) -- (REGc3.north) (REGc3.south) -- (REGi1.north) (REGi1.south) -- ($(REGc3.south |- REGf1.south)+(0, -0.2)$);

  \draw[-] (REGc3.Q) -- (REGc3.Q -| G1.west) (G1.east) -- (I2.west) (I2.east |- REGi2.D) -- (REGi2.D);
  \draw[-] (REGi1.Q) -- (C0R1.west) (C0R1.east |- REGc0r1.D) -- (REGc0r1.D);
  \draw[-] ($(C0R1.east)!0.5!(G1.west|-C0R1.east)$) node[branch] {} |- ($(G1.base west)!0.5!(G1.south west)$);

  \draw[dotted] (cc.north -| REGi2.north) -- (REGi2.north) (REGi2.south) -- (REGc0r1.north) (REGc0r1.south) -- ($(REGi2.south |- REGf1.south)+(0, -0.2)$);

  \draw[-] (REGi2.Q) -- (REGi2.Q -| C1.west);
  \draw[-] (REGc0r1.Q) -- (REGc0r1.Q -| C1.west);
  \draw[-] (REGcomb1.D -| C1.east) -- (REGcomb1.D);

  \draw[dotted] (cc.north -| REGcomb1.north) -- (REGcomb1.north) (REGcomb1.south) -- ($(REGcomb1.south |- REGf1.south)+(0, -0.2)$);

  \draw[-] (bc.west) -- (REGcomb1.Q);

  \draw let \p1 = (REGc0), \p2 = (cc) in node[coordinate] at (\x1, \y2) (cc0) {};
  \draw let \p1 = (REGc1), \p2 = (cc) in node[coordinate] at (\x1, \y2) (cc1) {};
  \draw let \p1 = (REGc2), \p2 = (cc) in node[coordinate] at (\x1, \y2) (cc2) {};
  \draw let \p1 = (REGc3), \p2 = (cc) in node[coordinate] at (\x1, \y2) (cc3) {};
  \draw let \p1 = (REGi2), \p2 = (cc) in node[coordinate] at (\x1, \y2) (cc4) {};
  \draw let \p1 = (REGcomb1), \p2 = (cc) in node[coordinate] at (\x1, \y2) (cc5) {};

  \node at ($(cc0)!0.5!(cc1)$) {1};
  \node at ($(cc1)!0.5!(cc2)$) {2};
  \node at ($(cc2)!0.5!(cc3)$) {3};
  \node at ($(cc3)!0.5!(cc4)$) {4};
  \node at ($(cc4)!0.5!(cc5)$) {5};

\end{tikzpicture}}
  \caption{Fully-unrolled deeply-pipelined decoder for a systematic (8, 5) polar code. Clock signals omitted for clarity. CC stands for \glsentrylong{cc}.}
  \label{fig:unrolled_deeply_arch_8_5}
\end{figure}
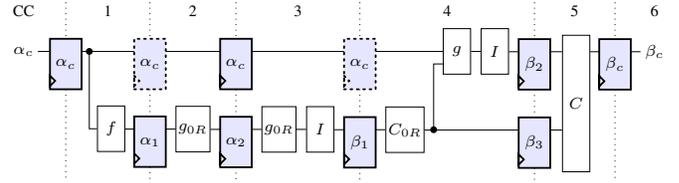

Unrolled decoder architectures provide extremely high decoding speeds. In an unrolled decoder architecture, each and every operation required is instantiated in hardware so that data can flow through the decoder with minimal control. 

\begin{table}[t]
\centering
\caption{Block types used in the unrolled decoders.}
\begin{tabular}{cl}
  \toprule
  \textbf{Name} & \textbf{Description}\\
  \midrule
  $f$ & Application of \eqref{eqn:sc:f}\vspace{2pt}\\
  $g$ & Application of \eqref{eqn:sc:g}\vspace{2pt}\\
  $g_{0R}$ & Application of \eqref{eqn:sc:g}, where $\mvec{\beta_l}$ is an all-zero vector\vspace{2pt}\\
  $I$ & Application of \eqref{eqn:sc:estu}, note that $i \notin \mathcal{F}$\vspace{2pt}\\
  $C$ & Application of \eqref{eqn:sc:combine}\vspace{2pt}\\
  $C_{0R}$ & Application of \eqref{eqn:sc:combine}, where $\mvec{\beta_l}$ is an all-zero vector\vspace{2pt}\\
  \bottomrule
\end{tabular}
\label{tab:block-types}
\end{table}

\autoref{fig:unrolled_deeply_arch_8_5} shows a fully-unrolled and deeply-pipelined decoder for the $(8,5)$ polar code illustrated in \autoref{fig:pc_8_5}. Data flows from left to right. The $\alpha$ and $\beta$ blocks illustrated in light blue are registers storing \glspl{llr} or bit estimates, respectively. White blocks are the functions described in \autoref{tab:block-types} and dotted registers are regular registers; they will be referred to when discussing partial pipelining in the following. As the $C_{0R}$ and $I$ blocks do not contain logic, i.e., they are equivalent to wires, they are inserted as preprocessing and postprocessing blocks, respectively. Among the registers, three are needed to retain the channel \glspl{llr}, denoted by $\alpha_c$ in the figure, during the $2^\text{nd}$, $3^\text{rd}$, and $4^{\text{th}}$ \glspl{cc}. Such unrolled architectures for polar decoders were described at length in \cite{Giard_TCASI_2016}.

\subsubsection*{Deeply Pipelined Vs Partially Pipelined}
In a deeply-pipelined architecture such as that illustrated in \autoref{fig:unrolled_deeply_arch_8_5}, a new frame is loaded into the decoder at every clock cycle. Therefore, a new estimated codeword is also output at each clock cycle. At any point in time, there are as many frames being decoded as there are pipeline stages. This leads to a very high throughput at the cost of high memory requirements.

Partial pipelining, on the other hand, allows to reduce the required area, at the cost of reducing the throughput, by removing redundant registers in parts of the pipeline where data remains unchanged over multiple clock cycles \cite{Giard_TCASI_2016}. Removing the dotted registers in \autoref{fig:unrolled_deeply_arch_8_5} results in an initiation interval of 2, meaning that at every second clock cycle, a new frame can be fed into the decoder and a new codeword is estimated. We note that the interval only affects the memory, not the computational elements, in the decoder.

\subsection{Functions as Look-up Tables}\label{sec:bg:lut}
A \gls{lut} is a type of memory that maps a set of input values to output values. It can provide quick access to precomputed values that would otherwise require complex calculations. In the same vein, \glspl{lut} can also be used to approximate mathematical functions.  
In this work, we use the expression \emph{\gls{lut}-based decoders} to refer to decoders in which arithmetic computations are replaced with look-up operations of integer-valued messages\cite{shah_coarsely_2019,Koike-Akino2019}. 
The reliability information is embedded in the integer valued messages stored in the \glspl{lut}. We carefully craft \glspl{lut} to replace the logic that would normally go into the $f$ and $g$ functions of \autoref{tab:block-types}, where the $g$ \glspl{lut} are also used for $g_{0R}$. Details regarding the design of these \glspl{lut} are given in the next section.

\section{Unrolled and Pipelined LUT-based Simplified Successive-Cancellation Decoding}\label{sec:gen}
\subsection{Look-up Table Generation}\label{sec:gen:lut}
The \glspl{lut} are designed using the \gls{ib} method\cite{tishby2000information}. The \gls{ib} framework clusters an observed quantity $y$ into its compressed form $t$ such that $\max_{p(t|y)}I(X;T)$. The random variable $X=x$ is the designated quantity of primary relevance, e.g., a bit value, while the random variable $T=t\in\mathcal{T}=\{ 0,1,\dots, |\mathcal{T}|-1\}$ is the compressed or quantized observation. The deterministic mapping $p(t|y)$ can be treated as a \gls{lut} with $y$ and $t$ as input and output, respectively. 

Each message $t$ is associated with a \gls{lut} $L_x(t)$ which can be computed from the distribution $p(x,t)$ that is provided by the \gls{ib} solution.
Further, the finite alphabet $\mathcal{T}$ for the quantized messages is chosen such that 
\begin{align*}
L_{x}\left(t = \sfrac{|\mathcal{T}|}{2} - 1 - j\right)= - L_{x}\left(t = \sfrac{|\mathcal{T}|}{2} + j\right),
\end{align*}
where $ j=0, \cdots , \frac{|\mathcal{T}|}{2} - 1$. 
Moreover, $L_x(t)>L_x(t^\prime)$ if $t>t^\prime$, $\forall \: t,t^\prime \in \mathcal{T}$. Such a cluster assignment enables hard decisions of the bit $x$ as the first half of the alphabet translates to negative LLRs while the second half translates to positive LLRs.

\begin{figure}
    \centering
    \resizebox{0.5\columnwidth}{!}{\begin{tikzpicture}[yscale=0.4, xscale=1, node distance=0.3cm, auto]
	
		\def \nodesize {0.5} 
		\def \VertDist {2.5}
		\def \HorDist {1}
		\def \HorDisplace {0.5}
		
		\node (U0) at (0-\HorDist,0) [draw, circle, minimum width = \nodesize cm]{$u_0$};
		\node (U1) at (0-\HorDist,0-\VertDist) [draw, circle, minimum width = \nodesize cm]{$u_1$};
		\node (Cn0) at (0,0) [draw, rectangle, minimum width = \nodesize cm,minimum height = \nodesize cm]{$+$};
		\node (Cn1) at (0,0-\VertDist) [draw, rectangle, minimum width = \nodesize cm,minimum height = \nodesize cm]{$=$};	
		\node (X0) at (\HorDist,0) [draw, circle, minimum width = \nodesize cm]{$x_0$};
		\node (X1) at (\HorDist,0-\VertDist) [draw, circle, minimum width = \nodesize cm]{$x_1$};
		\node (Cn2) at (2*\HorDist +\HorDisplace,0) [draw, rectangle, minimum width = \nodesize cm,minimum height = \nodesize cm]{$p(y|x)$};
		\node (Cn3) at (2*\HorDist +\HorDisplace,0-\VertDist) [draw, rectangle, minimum width = \nodesize cm,minimum height = \nodesize cm]{$p(y|x)$};	
		\node (Y0) at (3*\HorDist +2*\HorDisplace,0) [draw, circle, minimum width = \nodesize cm]{$y_0$};
		\node (Y1) at (3*\HorDist +2*\HorDisplace,0-\VertDist) [draw, circle, minimum width = \nodesize cm]{$y_1$};

		\draw[-] (U0) -- (Cn0);
		\draw[-] (U1) -- (Cn1);
		\draw[-] (Cn0) -- (X0);
		\draw[-] (Cn1) -- (X1);
		\draw[-] (U1) -- (Cn0);
		\draw[-] (X0) -- (Cn2) --(Y0);
		\draw[-] (X1) -- (Cn3) --(Y1);
		
	
	\end{tikzpicture}}
    \caption{Setup for generating decoding \glspl{lut} on a single building block}
    \label{fig:N2_LUT_setup}
\end{figure}
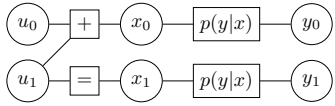

\autoref{fig:N2_LUT_setup} is used to illustrate the process of generating decoding \glspl{lut} for a polar code of length $N=2$.  In the figure, the uncoded bits $\mvec{u}$ are transformed into coded bits $\mvec{x}$ and received as $\mvec{y}$ after transmission through a  quantized \gls{awgn} channel with transition probabilities $p(y|x)$.  The \gls{awgn} channel quantizer is designed for a certain noise variance $\sigma_N^2$ using the \gls{ib} method with $y \in \mathcal{T}$.

With the quantized channel outputs $[y_0,y_1]$ at hand, the \glspl{lut} for decoding $u_0$ and $u_1$ are generated by quantizing their bit channels  using the \gls{ib} method. For decoding $u_0$, the output $\mvec{y_0}=[y_0,y_1]$ of its bit channel are quantized into $t_0\in\mathcal{T}$ such that $\max_{p(t_0|\mvec{y}_0)}I(U_0;T_0)$. The deterministic mapping $p(t_0|\mvec{y}_0)$ can serve as a \gls{lut} that replaces the $f$ function. It maps $\mvec{y_0}$ to $t_0$ which, in turn, can be used to decode $u_0$ as 
\begin{align}\label{eqn:ib:estu}
\hat{u}_i = \begin{cases}
0\text{,} & \text{when } t_i >\sfrac{|\mathcal{T}|}{2}\\
1\text{,} & \text{otherwise.}
\end{cases}
\end{align}
with $i = 0$. 
Similarly, the output $\mvec{y_1}=[y_0,y_1,u_0]$ of the bit channel of $u_1$ is compressed to $t_1 \in \mathcal{T}$   aiming $\max_{p(t_1|\mvec{y}_1)}I(U_1;T_1)$. The \gls{lut} $p(t_1|\mvec{y}_1)$ can then replace the $g$ function, where it is used to map $\mvec{y}_1$ to $t_1$ and decode $u_1$ using \eqref{eqn:ib:estu}. 

For a polar code of length $N > 2$, the decoding \glspl{lut} are obtained by integrating the aforementioned mutual information preserving quantization into its density evolution\cite{shah_coarsely_2019}. These decoding \glspl{lut} then replace \eqref{eqn:sc:f} and \eqref{eqn:sc:g} in the \gls{sc} decoding.
This is illustrated for the root node in the decoder tree of Fig.\,\ref{fig:sc-tree}, and \glspl{lut} $p(t_0|\mvec{y}_0)$ and $p(t_1|\mvec{y}_1)$ obtained in this section. In this case, the \gls{llr} carrying vectors $\mvec{\alpha}$ get replaced with vectors of integer valued messages $\mvec{t}$. With $N_v{=}8$, $\mvec{t}_v$ will carry the  quantized channel outputs $y_0,\dots,y_7$. The \gls{lut} $p(t_i| y_i,y_{i + \sfrac{N_v}{2}})$, i.e., the mapping $p(t_0|\mvec{y}_0)$ with labels adjusted for $N{=}8$, will be used to generate the updates $\mvec{t}_l$ for the left child of the root node. Similarly, the updates $\mvec{t}_r$ for the right child will be generated using the \gls{lut}  $p(t_i| y_i,y_{i + \sfrac{N_v}{2}},\beta_l[i])$, i.e., the mapping $p(t_1|\mvec{y}_1)$ with adjusted labels. Note that  a separate decoding \gls{lut} is used for each edge in the \gls{sc} decoding tree, i.e, $2N{-}N$ \glspl{lut} in total. At the leaf nodes, \eqref{eqn:ib:estu} can be used to estimate $\bm{\hat{u}}$ from $t_l$ and $t_r$.  The decoder that uses the \glspl{lut} designed using the \gls{ib} framework is henceforth referred to as the \gls{ib} decoder.

\subsection{Hardware-Efficient Look-up Tables}\label{sec:lut:variants}

The  \glspl{lut} in Section \ref{sec:gen:lut} are obtained from quantized density evolution of a polar code. 
Thus the $N{-}1$ distinct $f$-operation \glspl{lut} are designed according to the check-node update rule, i.e., the so-called box-plus operation. 
The $g$ \glspl{lut} are designed according to the variable-node update rule.
From a hardware point of view, efficient implementation of one, e.g., $f$, \gls{lut} might not be valid for a different $f$ \gls{lut} because the two will have different truth tables. However, it is beneficial to reduce the number of distinct \glspl{lut} so that the same hardware-efficient realization can be used for multiple \glspl{lut}.

In \cite{shah_MSIB_2023}, \gls{lut}-based polar decoders were constructed where the min-sum approximation is utilized for designing the $f$ \glspl{lut} while, like \cite{shah_coarsely_2019}, the $g$ \glspl{lut} are designed using the \gls{ib} method. It was shown in \cite{shah_MSIB_2023} that the effect of using the approximate min-sum rule for \gls{lut} design on the error-correction performance of the decoder is negligible. Such a decoder is referred to as an \emph{MS-IB} decoder.  

In an MS-IB decoder, all the $f$ \glspl{lut} have the same input/output relation. The number of distinct  $f$ \glspl{lut}, w.r.t. an \gls{ib} decoder, thus reduces from $N-1$ to 1. Most importantly, the min-sum based $f$ \gls{lut} can be realized as a min-sum of the integer valued message $t_a,t_b \in \mathcal{T}$. 
Recall (cf. Section \ref{sec:gen:lut}) that the integer messages embed \gls{llr} information. The \gls{msb} of the \gls{lut} inputs can be treated as the sign of the associated \gls{llr} with $\text{\gls{msb}}=0$ meaning negative \gls{llr} and vice versa. The remaining bits can be seen as the input-message magnitude associated to the \gls{llr}. The min-sum operation of the input integer messages can be expressed as:
\begin{align}\label{eqn:lut:minsum}
    t_o = f( t_a - \Delta, t_b - \Delta) + \Delta, 
\end{align}
where $\Delta= \frac{|\mathcal{T}|-1}{2}$, $t_o \in \mathcal{T}$ and $f(\cdot,\cdot)$ is computed using \eqref{eqn:sc:f}.

In comparison to \eqref{eqn:sc:f}, the argument $\Delta$ in \eqref{eqn:lut:minsum} is used for pre- and post-processing of the min-sum operation. 
From an implementation viewpoint, this is equivalent to inverting the magnitude carrying bits of any input message as well as the output message, if it falls in the first half of the finite alphabet $\mathcal{T}$. This extra logic can be discarded if a partially flipped finite alphabet is used. 
More precisely, all the integer messages are relabeled such that they belong to the \emph{relabeled} finite alphabet $\mathcal{T}_{re} = \{3,2,1,0,4,5,6,7 \}$ instead of  $\mathcal{T} = \{0,1,2,3,4,5,6,7 \}$ for $|\mathcal{T}|=8$.
Under the relabeled alphabet, the min-sum expression of \eqref{eqn:sc:f} can directly be applied to integer-valued messages $t_a^\prime,t_b^\prime\in\mathcal{T}_{re}$. A \gls{lut}-based MS-IB decoder that  uses $\mathcal{T}_{re}$ is referred to as \emph{re-MS-IB} decoder. Relabeling has no effect on error-correction performance.

\subsection{Unrolled-Architecture Generation}\label{sec:gen:unrolled}
The hardware implementations are generated using our software toolchain, first mentioned in \cite{Sarkis2014a} but significantly improved over the years to extend its functionality, including to generate hardware unrolled decoders \cite{Giard_TCASI_2016}. For this work, we further modified it in order to add support for substituting the $f$, $g$, $g_{0R}$ functions of \autoref{tab:block-types} by \glspl{lut}.

Our toolchain notably takes a polar code construction as well as a configuration file as input, and from that, it optimizes the decoder tree, e.g., going from the decoder tree of Fig.\,\ref{fig:sc-tree} to that of Fig.\,\ref{fig:ssc-tree}, and then generates the decoder. Among the configurable options are the code length, rate, the types of nodes that can be used, and the type of decoder. The type of nodes dictates the decoding algorithm. For this work, we generate hardware unrolled decoders in VHDL that implement the \gls{ssc} algorithm.

\section{Implementation and Results}\label{sec:impl}

In this section, we present implementation results for various \gls{lut}-based unrolled decoders. A fixed-point unrolled decoder is used for reference. Our decoders target a systematic $(128,64)$ polar code optimized for $\sfrac{E_b}{N_0}=3.0$\,dB using the method of Tal and Vardy \cite{Tal2011a}, and have an initiation interval of $10$ and a fixed latency of 86\,\glsentrylongpl{cc}. Without loss of generality, systematic coding is used as it offers better \gls{ber} performance than non-systematic coding at the cost of a negligible complexity increase. The quantization used for the fixed-point decoder was determined by way of simulation with bit-true models.
We denote quantization as $Q_i$.$Q_c$, where $Q_c$ is the total number of bits used to store a channel \gls{llr} and $Q_i$ is the total the number of bits used to store an internal \gls{llr}.
All \glspl{llr} use 2's complement representation.
All \gls{lut}-based decoders were designed for $\sfrac{E_b}{N_0}=3.0$\,dB and $|\mathcal{T}|=16$, i.e, 4-bit resolution.

\subsection{Functional Blocks in \gls{lut}-based Decoders}
This section presents how the block types of \autoref{tab:block-types} were adapated to the \gls{lut}-based decoders.

The circuits implementing the $C$ and $C_{0R}$ blocks are identical to the fixed-point decoder. The $I$ blocks are similar to those of the fixed-point decoder, with an added inverter per bit. It corresponds to a series of inverters that take the \gls{msb} of the soft messages $\mvec{\alpha}$ or the integer messages $\mvec{t}$ as input. 

The $g$ and $g_{0R}$ blocks are realized by the synthesis tools as logic circuits according to the truth table of corresponding \glspl{lut} in the three \gls{lut}-based decoders. In the \gls{ib} decoder, the ${f}$ blocks are realized in the same way.

\begin{figure}[t]
  \centering
  \resizebox{\columnwidth}{!}{\tikzset{
  branch/.style={fill,shape=circle,minimum size=4pt,inner sep=0pt},
}

\begin{tikzpicture}[circuit logic US,font=\normalsize,scale=2]

  \node[xor gate] at (0, 0) (sign) {};

  \node[shape=mux2, anchor=south, yscale=-1] (mout) at ($(sign.south)+(1.3,-0.4)$) {};
  \draw[-] (sign.output -| mout.sel) node[branch] {} -- (mout.sel);
  \node[anchor=west] (mouti0) at ($(mout.in0)+(-0.05, 0)$) {1};
  \node[anchor=west] (mouti1) at ($(mout.in1)+(-0.05, 0)$) {0};
  \node[not gate, anchor=west, scale=0.6] (nsign) at ($(sign.output)+(1.0,0)$) {};
  \draw[-] (sign.output) -- (nsign.input);

  \node[shape=mux2, anchor=south] (mamp) at (mout.in0 -| sign) {};
  \node[anchor=west] (mampi0) at ($(mamp.in0)+(-0.05, 0)$) {0};
  \node[anchor=west] (mampi1) at ($(mamp.in1)+(-0.05, 0)$) {1};
  \node[not gate, scale=0.6] (namp) at ($(mamp.out)!0.5!(mout.in1)$) {};
  \draw[-, very thick] (mamp.out) -- (namp.input) (namp.output) -- (mout.in1);
  \draw[-, very thick] ($(mamp.out)!0.5!(namp.input)$) node[branch] {} |- (mout.in0);

  \node[shape=mux2,yscale=-1] (maamp) at ($(mamp.in0)+(-1.0, 0)$) {};
  \node[anchor=west] (maampi0) at ($(maamp.in0)+(-0.05, 0)$) {1};
  \node[anchor=west] (maampi1) at ($(maamp.in1)+(-0.05, 0)$) {0};
  \draw[-, very thick] (maamp.out) -- (mamp.in0);

  \node[shape=mux2,anchor=south,yscale=-1] (mbamp) at ($(maamp.north)+(0, -0.3)$) {};
  \node[anchor=west] (mbampi0) at ($(mbamp.in0)+(-0.05, 0)$) {1};
  \node[anchor=west] (mbampi1) at ($(mbamp.in1)+(-0.05, 0)$) {0};

  \node[circle,draw,anchor=south west] at ($(mbamp.out -| sign.input 1)+(0,0)$) (cmp) {$>$};
  \draw[-] (cmp.east) -| (mamp.sel);
  \node (aa) at ($(maamp.out)!0.5!(mamp.in0)$) {};
  \draw[-,very thick] (aa.center) node[branch]{} |- (cmp.north west);
  \draw[-,very thick] (mbamp.out) -- (cmp.south west);

  \draw[-, very thick] ($(mbamp.out)!0.4!(cmp.south west)$) node[branch] {} |- (mamp.in1);

  \node[anchor=west,align=center](inAa) at ($(maamp.in1)+(-1.8, 0)$) {$t_{a,2}$ \\[-0.3em]$t_{a,1}$\\[-0.3em]$t_{a,0}$};
  \node[not gate, scale=0.6] (naamp) at ($(inAa.east)!0.6!(maamp.in1)$) {};
  \draw[-, very thick] (inAa.east) -- (naamp.input) (naamp.output) -- (maamp.in1);
  \draw[-, very thick] ($(inAa.east)!0.7!(naamp.input)$) node[branch] {} |- (maamp.in0);

  \node[anchor=west,align=center](inBa) at (mbamp.in1 -| inAa.west) {$t_{b,2}$ \\[-0.3em]$t_{b,1}$ \\[-0.3em]$t_{b,0}$};
  \node[not gate, scale=0.6] (nbamp) at ($(inBa.east)!0.6!(mbamp.in1)$) {};
  \draw[-, very thick] (inBa.east) -- (nbamp.input) (nbamp.output) -- (mbamp.in1);
  \draw[-, very thick] ($(inBa.east)!0.7!(nbamp.input)$) node[branch] {} |- (mbamp.in0);

  \node[anchor=west](inAs) at ($(sign.input 1 -| inAa.west)+(-.3,0)$) {$t_{a,3}$};
  \draw[-] (inAs.east) -- (sign.input 1);
  \node(inBs) at (sign.input 2 -| inAs) {$t_{b,3}$};
  \draw[-] (inBs.east) -- (sign.input 2);

  \draw[-] (inAs -| maamp.sel) node[branch] {} -- (maamp.sel);
  \node (bs) at ($(inBs.east)+(0.6, 0)$) {};
  \draw[-] (bs.center) node[branch] {} |- ($(bs |- mout.out)+(0, -0.1)$) -| (mbamp.sel);

  \node[anchor=west](outOs) at ($(nsign.output)+(0.2,0)$) {$t_{o,3}$};
  \node[anchor=west, align=center](outOa) at (mout.out -| outOs.west) {$t_{o,2}$ \\[-0.3em]$t_{o,1}$ \\[-0.3em]$t_{o,0}$};
    
  \draw[-] (nsign.output) -- (outOs.west);
  \draw[-,very thick] (mout.out) -- (outOa.west);
\end{tikzpicture}}
  \caption{min-sum for $\mathcal{T}$ in the MS-IB decoder.}
  \label{fig:minf2}
\end{figure}
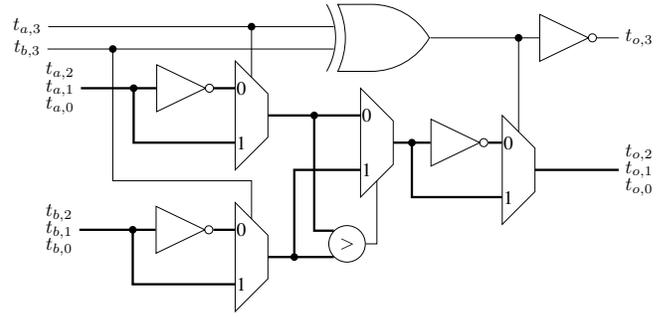
\autoref{fig:minf2} illustrates the $f$ block circuit for the MS-IB decoder. The binary form of a 4-bit message $t_{x}$ is denoted as $[t_{x,3},t_{x,2},t_{x,1},t_{x,0}]$. Thick lines indicate vectors of bits. By slight abuse of symbolism, the inverters with vectors as input and output carry out bitwise inversions.

\begin{figure}[t]
  \centering
  \resizebox{0.6\columnwidth}{!}{\tikzset{
  branch/.style={fill,shape=circle,minimum size=4pt,inner sep=0pt},
}

\begin{tikzpicture}[circuit logic US,font=\normalsize,scale=2]

  \node[xnor gate] at (0, 0) (sign) {};
  \node[anchor=east](inAs) at ($(sign.input 1)+(-1.25, 0)$) {$t_{a,3}$};
  \node[anchor=east](inBs) at ($(sign.input 2)+(-1.25, 0)$) {$t_{b,3}$};

  \draw[-] (inAs.east) -- (sign.input 1);
  \draw[-] (inBs.east) -- (sign.input 2);

  \node[shape=mux2, anchor=north] (m0) at ($(sign.south east)+(0.4,-0.1)$) {};
  \node[anchor=west] (m0i0) at ($(m0.in0)+(-0.05, 0)$) {0};
  \node[anchor=west] (m0i1) at ($(m0.in1)+(-0.05, 0)$) {1};

  \node[anchor=west,align=center](inAa) at ($(m0.in0 -| inAs.west)+(.3,0)$) {$t_{a,2}$ \\[-0.3em]$t_{a,1}$ \\[-0.3em]$t_{a,0}$};
  \node[anchor=west,align=center](inBa) at ($(m0.in1 -| inAa.west)+(.3,0)$) {$t_{b,2}$ \\[-0.3em]$t_{b,1}$ \\[-0.3em]$t_{b,0}$};
  \draw[-,very thick] (inAa.east) -- (m0.in0);
  \draw[-,very thick] (inBa.east) -- (m0.in1);

  \node[circle,draw] at ($(m0.south)+(-0.4, -0.2)$) (cmp) {$>$};
  \draw[-] (cmp.east) -| (m0.sel);
  \draw[-,very thick] ($(inAa.east)!0.5!(m0.in0)$) node[branch]{} |- (cmp.north west);
  \draw[-,very thick] ($(inBa.east)!0.25!(m0.in1)$) node[branch]{} |- (cmp.south west);

  \node[anchor=west](outOs) at ($(sign.output)+(0.5,0)$) {$t_{o,3}$};
  \node[anchor=west,align=center](outOa) at (m0.out -| outOs.west) {$t_{o,2}$ \\[-0.3em]$t_{o,1}$ \\[-0.3em]$t_{o,0}$};

  \draw[-] (sign.output) -- (outOs.west);
  \draw[-,very thick] (m0.out) -- (outOa.west);
\end{tikzpicture}}
  \caption{min-sum for $\mathcal{T}_{re}$ in the re-MS-IB decoder.}
  \label{fig:minf3}
\end{figure}
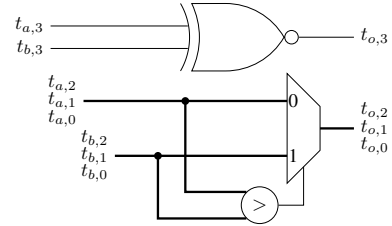

\autoref{fig:minf3} illustrates the $f$ block circuit for the re-MS-IB variant. Comparing Figs.\,\ref{fig:minf2} and \ref{fig:minf3}, it can be seen that the critical path for the re-MS-IB variant of $f$ is much shorter than that of MS-IB. It consists of a single multiplexer as opposed to 3 multiplexers and 2 inverters. Furthermore, the amount of resources required is much less, as will be reflected in the results of Section\,\ref{sec:impl:cmp}.

\subsection{Error-correction Performance and Impact of Quantization}\label{sec:impl:ec-perf}
\begin{figure}
  \centering
  \begin{tikzpicture}[spy using outlines={circle, magnification=2.75, connect spies}]

  \pgfplotsset{
    grid style = {
      dash pattern = on 0.05mm off 1mm,
      line cap = round,
      black,
      line width = 0.5pt
    },
    label style = {font=\fontsize{10pt}{7.2}\selectfont},
    tick label style = {font=\fontsize{8pt}{7.2}\selectfont}
  }

  \begin{semilogyaxis}[%
    xlabel=$E_b/N_0$ (dB),%
    xlabel style={yshift=0.5em},%
    xtick={0, 1, ..., 5},%
    xmin=0, xmax=5,
    ymin=5e-5,
    ylabel=FER, ylabel style={yshift=-0.6em},%
    width=0.52\columnwidth, height=6.6cm, grid=major,%
    legend style={
      anchor={center},
      cells={anchor=west},
      column sep= 2mm,
      font=\fontsize{8pt}{7.2}\selectfont
    },
    legend to name=ec-perf-legend,
    legend columns=3,
    mark size=2.0pt,
    mark options=solid]

    \addlegendimage{empty legend}
    \addlegendentry[anchor=east]{Floating point:}
    \addplot[color=black,solid,thick] table[x=ebn0_db,y=FER] {data/n128_awgn_s0.708.perf.trex.float.csv};
    \addlegendentry{}
    \addlegendimage{empty legend}
    \addlegendentry{}

    \addlegendimage{empty legend}
    \addlegendentry[anchor=east]{Fixed point:}
    
    \addplot[color=blue,thick,dashed,mark=o,mark options=solid] table[x=ebn0_db,y=FER] {data/n128_awgn_s0.708.perf.trex.fixed.5.4.0.csv};
    \addlegendentry{$Q_i.Q_c=5.4$}

    \addplot[color=orange,thick,dashed,mark=square,mark options=solid] table[x=ebn0_db,y=FER] {data/n128_awgn_s0.708.perf.trex.fixed.4.4.0.csv};
    \addlegendentry{$Q_i.Q_c=4.4$}

    \addlegendimage{empty legend}
    \addlegendentry[anchor=east]{LUTs:}

    \addplot[color=red,thick,dotted,mark=+,mark options=solid] table[x=ebn0_db,y=FER] {data/n128_awgn_s0.708.perf.HW-Unrolled-LUT.csv};
    \addlegendentry{IB}

    \addplot[color=teal,thick,dotted,mark=asterisk,mark options=solid] table[x=EbN0dB,y=FER] {data/n128_awgn_s0.708.perf.HW-Unrolled-minLUT-relabel-SSC.csv};
    \addlegendentry{MS-IB/re-MS-IB}

    \coordinate (spypoint) at (axis cs:4.35,1e-3); 
    \coordinate (magnifyglass) at (axis cs:2.2,5e-4); 
  
  \end{semilogyaxis}

  \spy [gray, size=1.5cm] on (spypoint) in node[fill=white] at (magnifyglass);
\end{tikzpicture}\hspace{-5pt}
\begin{tikzpicture}[spy using outlines={circle, magnification=2.75, connect spies}]

  \pgfplotsset{
    grid style = {
      dash pattern = on 0.05mm off 1mm,
      line cap = round,
      black,
      line width = 0.5pt
    },
    label style = {font=\fontsize{10pt}{7.2}\selectfont},
    tick label style = {font=\fontsize{8pt}{7.2}\selectfont}
  }

  \begin{semilogyaxis}[%
    xlabel=$E_b/N_0$ (dB),%
    xlabel style={yshift=0.5em},%
    xtick={0, 1, ..., 5},%
    xmin=0, xmax=5,
    ymin=5e-6,
    ylabel=BER, ylabel style={yshift=-0.6em},%
    width=0.52\columnwidth, height=6.6cm, grid=major,%
    mark size=2.0pt,
    mark options=solid]

    \addplot[color=black,thick,solid] table[x=ebn0_db,y=BER] {data/n128_awgn_s0.708.perf.trex.float.csv};

    \addplot[color=blue,thick,dashed,mark=o] table[x=ebn0_db,y=BER] {data/n128_awgn_s0.708.perf.trex.fixed.5.4.0.csv};

    \addplot[color=orange,thick,dashed,mark=square,mark options=solid] table[x=ebn0_db,y=BER] {data/n128_awgn_s0.708.perf.trex.fixed.4.4.0.csv};

    \addplot[color=red,thick,dotted,mark=+] table[x=ebn0_db,y=BER] {data/n128_awgn_s0.708.perf.HW-Unrolled-LUT.csv};

    \addplot[color=teal,thick,dotted,mark=asterisk,mark options=solid] table[x=EbN0dB,y=BER] {data/n128_awgn_s0.708.perf.HW-Unrolled-minLUT-relabel-SSC.csv};

    \coordinate (spypoint) at (axis cs:4.35,1e-4); 
    \coordinate (magnifyglass) at (axis cs:2.2,5e-5); 
  \end{semilogyaxis}
  \spy [gray, size=1.5cm] on (spypoint) in node[fill=white] at (magnifyglass);
\end{tikzpicture}
\\
\ref{ec-perf-legend}
  \caption{Error-correction performance of the systematic $(128, 64)$ polar code.}
  \label{fig:ecc-cmp}
\end{figure}
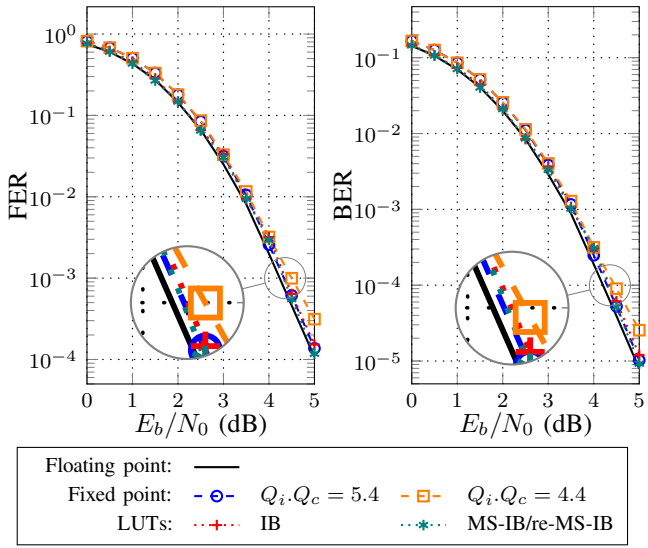

\autoref{fig:ecc-cmp} shows the error-correction performance of the systematic $(128, 64)$ polar code our unrolled decoders support, modulated with \gls{bpsk} and transmitted over an \gls{awgn} channel.
Fixed-point results for $Q_i$.$Q_c=5.4$ and $Q_i$.$Q_c=4.4$ are presented as well as results for all \gls{lut}-based variants.
Floating-point results are also included for reference.
The $Q_i$.$Q_c=5.4$ fixed-point decoder and the MS-IB and re-MS-IB \gls{lut}-based variants are shown to have a coding loss of under 0.1\,dB at a \gls{fer} of $10^{-3}$ or at a \gls{ber} of $10^{-4}$ compared to the floating-point representation.
Meanwhile, the coding loss of the \gls{ib} \gls{lut}-based decoder is little under 0.13\,dB at the same \gls{ber} and \gls{fer} values. The \gls{lut}-based decoders initially have a smaller coding loss than the $Q_i$.$Q_c=5.4$ fixed-point decoder, but eventually come to match as the channel noise decreases.

\subsection{Comparison of the Unrolled Decoders}\label{sec:impl:cmp}
\begin{table}
  \centering
  \caption{Comparison of unrolled decoders for a systematic $(128,64)$ polar code. All decoders have an initiation interval of $10$.}
  \begin{tabular}{l c ccc}
    \toprule
                 & \bf Fixed point& \multicolumn{3}{c}{\bf \gls{lut} based} \\
    \midrule
    \textbf{Variant}      & - & IB & MS-IB & re-MS-IB \\
    \textbf{Area} (mm$^2$)&          0.090 & 0.254 & 0.218 & 0.069 \\ 
    \textbf{Frequency} (GHz)&        1.47 &  1.38 &  1.40 & 1.51 \\
    \textbf{Latency} (ns)&           58.6 &  62.2 &  61.3 & 56.8 \\ 
    \textbf{Info. T/P} (Gbps)&       9.40 &  8.85 &  8.98 & 9.68 \\ 
    \textbf{Area Eff.} (Gbps/mm$^2$)& 104.3 & 34.9 &  41.2 & 140.0 \\ 
    \bottomrule
  \end{tabular}
  \label{tab:cmp_asic}
\end{table}

\autoref{tab:cmp_asic} shows synthesis results using a 28\,nm FD-SOI CMOS technology from ST Microelectronics. All decoders were synthesized to target a clock frequency of 1.5\,GHz. The first column is for the $Q_i$.$Q_c=5.4$ fixed-point decoder and the last three for the 4-bit \gls{lut}-based decoders. 

We observe that our first and relatively naive \gls{lut}-based implementation (\gls{ib}) unfortunately has 182\% higher area and 6\% lower throughput than the baseline fixed-point decoder. The situation improves when using the $f$ block of \autoref{fig:minf2}. The MS-IB decoder requires 142\% higher area for a 4\% lower throughput compared to the fixed-point decoder. Finally, significant gains with respect to the baseline decoder are observed when applying the relabeling described in Section~\ref{sec:lut:variants}, where the resulting decoder, re-MS-IB, is 23\% smaller and 3\% faster than the fixed-point decoder, leading to a 35\% better area efficiency.

\section{Conclusion}\label{sec:conclusion}
In this work, we showed that replacing blocks of logic by \glspl{lut} in an unrolled decoder for polar codes may not necessarily lead to gains in terms of key performance metrics. We presented three variants of \gls{lut}-based decoders and compared them against a regular fixed-point decoder. We used a short systematic polar code to illustrate. Beyond carefully crafting the \glspl{lut}, the key ingredient to obtain good performance has been to determine \gls{lut} realizations having  truth tables that lead to efficient implementation. This was achieved by deploying the min-sum approximate \gls{lut} design together with appropriate relabeling of the inputs and output of \glspl{lut}. As a result, the third variant of the \gls{lut}-based decoders (re-MS-IB) was shown to outperform the baseline fixed-point decoder on all metrics, notably offering 35\% better area efficiency with similar or better error-correction performance.

\section*{Acknowledgement}
The authors would like to thank Marzieh Hashemipour Nazari (Eindhoven University of Technology) for providing the ASIC synthesis results. The work of Pascal Giard is supported by NSERC's Discovery Grant \#651824.

\bibliographystyle{IEEEtran}
\bibliography{IEEEabrv,ConfAbrv,references}

\end{document}